\documentclass[sigconf]{acmart}

\usepackage{url}
\usepackage{amsmath}
\usepackage{makecell}
\usepackage{float}
\usepackage{graphicx}

\usepackage{listings}
\usepackage{bera}

\newcommand{\SOURCECODE}{\url{https://github.com/endgameinc/ember}}

\begin{document}

\title{EMBER: An Open Dataset for Training Static PE Malware Machine Learning Models}

	\author{Hyrum S.~Anderson}
	\affiliation{ \institution{ Endgame, Inc.} }
	\email{hyrum@endgame.com}

	\author{Phil Roth}
    \affiliation{ \institution{ Endgame, Inc.} }
    \email{proth@endgame.com}

\begin{abstract}
This paper describes EMBER: a labeled benchmark dataset for training machine learning models to statically detect malicious Windows portable executable files.  The dataset includes features extracted from 1.1M binary files: 900K training samples (300K malicious, 300K benign, 300K unlabeled) and 200K test samples (100K malicious, 100K benign).  To accompany the dataset, we also release open source code for extracting features from additional binaries so that additional sample features can be appended to the dataset. This dataset fills a void in the information security machine learning community: a benign/malicious dataset that is large, open and general enough to cover several interesting use cases. We enumerate several use cases that we considered when structuring the dataset. Additionally, we demonstrate one use case wherein we compare a baseline gradient boosted decision tree model trained using \texttt{LightGBM} with default settings to MalConv, a recently published end-to-end (featureless) deep learning model for malware detection. Results show that even without hyper-parameter optimization, the baseline EMBER model outperforms MalConv. The authors hope that the dataset, code and baseline model provided by EMBER will help invigorate machine learning research for malware detection, in much the same way that benchmark datasets have advanced computer vision research.
\end{abstract}


\maketitle

\section{Introduction}
\label{sec:intro}

Machine learning can be an attractive tool for either a primary detection capability or supplementary detection heuristics.  Supervised learning models automatically exploit complex relationships between file attributes in training data that are discriminating between malicious and benign samples. 
Furthermore, properly regularized machine learning models generalize to new samples whose features and labels follow a similar distribution to the training data. 

However, malware detection using machine learning has not received nearly the same attention in the open research community as other applications, where rich benchmark datasets exist.   These include handwritten digit classification (e.g., MNIST \cite{MNIST}), image labeling (e.g., CIFAR \cite{CIFAR} or ImageNet \cite{imagenet}), traffic sign detection \cite{trafficsigns}, speech recognition (e.g., TIMIT \cite{TIMIT}), sentiment analysis (e.g., Sentiment140 \cite{sentiment140}), and a host of other datasets suitable for training models to mimic human perception and cognition tasks.  The challenges to releasing a benchmark dataset for malware detection are many, and may include the following.
\begin{itemize}
\item \textit{Legal restrictions.} Malicious binaries are shared generously through sites like VirusShare \cite{virusshare} and VX Heaven \cite{vxheaven}, but benign binaries are often protected by copyright laws that prevent sharing.  Both benign and malicious binaries may be obtained at volume for internal use through for-pay services such as VirusTotal \cite{virustotal}, but subsequent sharing is prohibited.

\item \textit{Labeling challenges.}  Unlike images, text and speech---which may be labeled relatively quickly, and in many cases by a non-expert \cite{buhrmester2011amazon}---determining whether a binary file is malicious or benign can be a time-consuming process for even the well-trained.  The work of labeling may be automated via antimalware scanners that codify much of this human expertise, but the results may be proprietary or otherwise protected.  Aggregating services like VirusTotal specifically restrict the public sharing of vendor antimalware labels \cite{virustotal}.

\item \textit{Security liability and precautions.} There may be risk in promoting a large dataset that includes malicious binaries to a general non-infosec audience not accustomed to taking appropriate precautions such as sandboxed hosting.
\end{itemize}

We address these issues with the release of the Endgame Malware BEnchmark for Research (EMBER) dataset\footnote{Data and code available at \SOURCECODE}, extracted from a large corpus of Windows portable executable (PE) malicious and benign files.  This allows free dissemination of both malicious and benign entities without legal or security concerns.  Samples are released together with the sha256 hash of the original file, and a label to denote whether the file is deemed to be malicious or benign.  The pre-selection of features naturally limits the flexibility of researchers in comparing feature sets.  This is somewhat ameliorated by our release of open source code to compute the PE features for feature comparison studies. The lack of raw binaries also precludes experiments using featureless deep-learning malware detectors (e.g., \cite{raff2017malware}).  However, we hope that by releasing the sha256 hashes, feature extraction source code, as well as a high-performing baseline classifier computed from a subset of the features, this dataset and model codebase will still become a relevant baseline for machine learning malware detection research, and to which featureless deep learning studies may compare.  We demonstrate such a comparison in Section \ref{sec:results}.

We begin in Section \ref{sec:background} with relevant background about the PE file format, as well as a summary of related datasets and approaches for static malware classification.  In Section \ref{sec:method}, we describe the dataset and our methodology for its format.  We demonstrate the efficacy of our baseline model trained on this dataset in Section \ref{sec:results}. Source code and data can be found at \SOURCECODE.

\section{Background}
\label{sec:background}
We summarize important context in the portable executable (PE) file format in Section~\ref{sec:pe}.  In Section~\ref{sec:staticpe}, we review related work in feature extraction for classifying malware using machine learning.  Finally, we summarize other relevant static malware datasets in Section~\ref{sec:datasets}.  

\begin{figure}[t]
    \centering
    \includegraphics[width=3.6in]{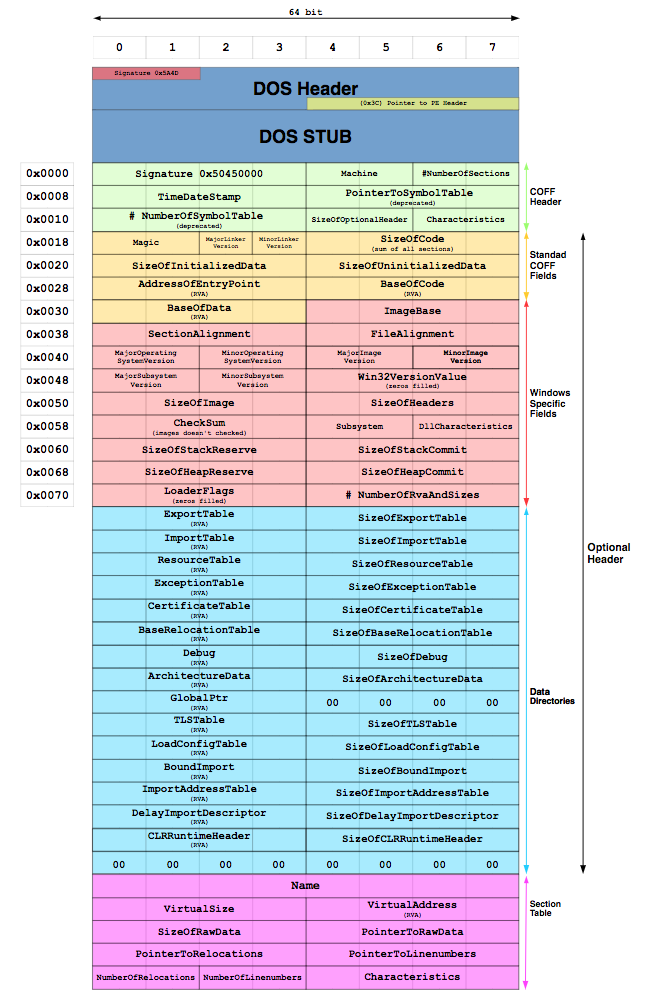}
    \caption{The 32-bit PE file structure. Creative commons image courtesy \cite{PEformat}.}
    \label{fig:pe_header}
\end{figure}

\subsection{PE File Format} 
\label{sec:pe}
The PE file format describes the predominant executable format for Microsoft Windows operating systems, and includes executables, dynamically-linked libraries (DLLs), and FON font files. The format is currently supported on Intel, AMD and variants of ARM instruction set architectures.

The file format is arranged with a number of standard headers (see Fig. \ref{fig:pe_header} for PE-32 format), followed by one or more sections \cite{pietrek2002inside}.  Headers include the Common Object File Format (COFF) file header that contains important information such as the type of machine for which the file is intended, the nature of the file (DLL, EXE, OBJ), the number of sections, the number of symbols, etc.  The optional header identifies the linker version, the size of the code, the size of initialized and uninitialized data, the address of the entry point, etc.  Data directories within the optional header provide pointers to the sections that follow it.  This includes tables for exports, imports, resources, exceptions, debug information, certificate information, and relocation tables.  As such, it provides a useful summary of the contents of an executable \cite{shafiq2009pe}. Finally, the section table outlines the name, offset and size of each section in the PE file.

PE sections contain code and initialized data that the Windows loader is to map into executable or readable/writeable memory pages, respectively, as well as imports, exports and resources defined by the file.  Each section contains a header that specifies the size and address.  An import address table instructs the loader which functions to statically import.  A resources section may contain resources such as required for user interfaces: cursors, fonts, bitmaps, icons, menus, etc.  A basic PE file would normally contain a \texttt{.text} code section and one or more data sections (\texttt{.data}, \texttt{.rdata} or \texttt{.bss}).  Relocation tables are typically stored in a \text{.reloc} section, used by the Windows loader to reassign a base address from the executable's preferred base.  A \texttt{.tls} section contains special thread local storage (TLS) structure for storing thread-specific local variables, which has been exploited to redirect the entry point of an executable to first check if a debugger or other analysis tool are being run \cite{brand2010malware}.  Section names are arbitrary from the perspective of the Windows loader, but specific names have been adopted by precedent and are overwhelmingly common.  Packers may create new sections, for example, the UPX packer creates \texttt{UPX1} to house packed data and an empty section \texttt{UPX0} that reserves an address range for runtime unpacking \cite{devi2012pe}.
\pagebreak[3]

\subsection{Static PE Malware Detection} \label{sec:staticpe}
Static malware detection attempts to classify samples as malicious or benign without executing them, in contrast to dynamic malware detection which detects malware based on its runtime behavior including time-dependent sequences of system calls for analysis~\cite{dahl2013large,pascanu2015malware,athiwaratkun2017malware}. Although static detection is well-known to be undecidable in general~\cite{cohen1987computer}, it is an important protection layer in a security suite because when successful, it allows malicious files to be detected prior to execution. 

Machine learning-based static PE malware detectors have been used since at least 2001~\cite{schultz2001data}, and owing largely to the structured file format and backwards-compatibility requirements, many concepts remain surprisingly similar in subsequent works~\cite{kolter2004learning, shafiq2009framework, raman2012selecting, dahl2013large, saxe2015deep}.  Schultz et al.~\cite{schultz2001data} assembled a dataset and generated labels by running through a McAfee virus scanner.  PE files were represented by features that included imported functions, strings and byte sequences.  Various machine learning models were trained and validated on a holdout set. Models included rules induced from RIPPER~\cite{cohen1995fast}, na\"{i}ve Bayes and an ensemble classifier. Kolter et al.~\cite{kolter2004learning} extended this approach by including byte-level N-grams, and employed techniques from natural language processing, including tf-idf weighting of strings.  Shafiq et al.~\cite{shafiq2009framework} proposed using just seven features from the PE header (described in Section \ref{sec:datasets}), motivated by the fact that most malware samples in their study typically exhibited those elements.  Saxe and Berlin leveraged novel two dimensional byte entropy histograms that is fed into a multi-layer neural network for classification~\cite{saxe2015deep}.  

Recent advances in end-to-end deep learning have dramatically improved the state of the art especially in object classification, machine translation and speech recognition.  In many of these approaches, raw images, text or speech waveforms are used as input to a machine learning model which infers the most useful feature representation for the task at hand.  However, despite successes in other domains hand-crafted features apparently still represent the state of the art for malware detection in published literature.  The state of the art may change to end-to-end deep learning in the ensuing months or years, but hand-crafted features derived from parsing the PE file may continue to be relevant indefinitely because of the structured format.  A recent example of end-to-end deep learning for malware classification is discussed in \cite{raff2017malware}, which we re-implement and compare to our baseline model in Section \ref{sec:results}.

\subsection{Malicious and benign datasets}
\label{sec:datasets}
PE-Miner aimed to produce a machine-learning based malware detector that exceeded 99\% true positive rate (TPR) at less than a 1\% false positive rate (FPR), with a runtime comparable to signature-based scanners of the day \cite{shafiq2009pe}.  It was trained on a dataset of $1,447$ benign files on the operating system (never published), $10,339$ malicious PE files from VX Heaven \cite{vxheaven}, and $5,586$ malicious PE files from Malfease.  PE-Miner uses 189 features that include a binary indicator for specific DLLs referenced, the sizes of various sections, summary information from the COFF section, a summary of the resource table, etc.  Unfortunately, many of the features were not disclosed publicly, some being deemed sensitive and protected under NDA \cite{seymour2016howto}.  Several model types were evaluated on the dataset, and of those, it was discovered that the J48 decision tree algorithm provided the best performance.  Notably, although many papers cite this work as one of the first performant (both speed and TP/FP rates) non-signature-based methods, the lack of public dataset has resulted no real comparative study.

Shortly following, the Adobe Malware Classifier aimed to produce a malware classifier from only seven features\footnote{The feature set we release in EMBER contains all of the information required to recreate features for the Adobe Malware Classifier.}: debug size, image version, relative virtual address of the import address table, export size, resource size, virtual size of the second section, and the total number of sections \cite{raman2012selecting}. A decision tree algorithm was trained and the resulting classifier released as a freely available tool\footnote{\url{https://sourceforge.net/adobe/malclassifier/wiki/Home/}}.  It has been suggested, however, that since the benign dataset largely comprised of Windows binaries, the resulting model is strongly biased towards a non-Windows vs. Windows rather than a malicious vs. benign problem \cite{seymour2016howto}. Indeed, in evaluating the pretrained model on the EMBER test set, we observe extremely large false positive rates and low detection rates (see Section \ref{sec:results}). Unfortunately, the dataset consisting of about 100K malicious files and 16K benign files was never released for comparative research.  

In contrast, the Microsoft Malware Classification Challenge concluded in April 2015 \cite{ronen2018microsoft}.  The dataset included a large dataset of 500MB, consisting of disassembly and bytecode of around 20K malicious samples from nine families.   The largest family consisted of features from 3K samples (\texttt{Kelihos} backdoor), while the smallest family included only 42 samples (\texttt{Simda} backdoor).  Since the conclusion of the competition, more than 50 research papers and theses cited the dataset.  A contributional summary of many of these works are tabulated in \cite{ronen2018microsoft}.  Unfortunately, the disassembly features are specific to IDA Pro disassembler (not easily reproducible), and the dataset contains no benign files.

Malware sharing services like VXHeaven provide an ample supply of malicious binaries \cite{vxheaven}.  VirusTotal can be mined for supposed benign files using heuristics about the number of detections of vendor participants \cite{virustotal}.  However, large-scale file access rates in VirusTotal require a payed subscription.  Regardless, an agreed-upon set of malicious and benign files for machine learning benchmark purposes is so far non-existent.

\pagebreak[3]
\section{Data Description}
\label{sec:method}
\begin{figure}[pt!]
\scriptsize
\begin{lstlisting}
  "|\textbf{sha256}|": "000185977be72c8b007ac347b73ceb1ba3e5e4dae4fe98d4f2ea92250f7f580e",
  "|\textbf{appeared}|": "2017-01",
  "|\textbf{label}|": -1,
  "|\textbf{general}|": {
    "|\textbf{file\_size}|": 33334,
    "|\textbf{vsize}|": 45056,
    "|\textbf{has\_debug}|": 0,
    "|\textbf{exports}|": 0,
    "|\textbf{imports}|": 41,
    "|\textbf{has\_relocations}|": 1,
    "|\textbf{has\_resources}|": 0,
    "|\textbf{has\_signature}|": 0,
    "|\textbf{has\_tls}|": 0,
    "|\textbf{symbols}|": 0
  },
  "|\textbf{header}|": {
    "|\textbf{coff}|": {
      "|\textbf{timestamp}|": 1365446976,
      "|\textbf{machine}|": "I386",
      "|\textbf{characteristics}|": [ "LARGE_ADDRESS_AWARE", ..., "EXECUTABLE_IMAGE" ]
    },
    "|\textbf{optional}|": {
      "|\textbf{subsystem}|": "WINDOWS_CUI",
      "|\textbf{dll\_characteristics}|": [ "DYNAMIC_BASE", ..., "TERMINAL_SERVER_AWARE" ],
      "|\textbf{magic}|": "PE32",
      "|\textbf{major\_image\_version}|": 1,
      "|\textbf{minor\_image\_version}|": 2,
      "|\textbf{major\_linker\_version}|": 11,
      "|\textbf{minor\_linker\_version}|": 0,
      "|\textbf{major\_operating\_system\_version}|": 6,
      "|\textbf{minor\_operating\_system\_version}|": 0,
      "|\textbf{major\_subsystem\_version}|": 6,
      "|\textbf{minor\_subsystem\_version}|": 0,
      "|\textbf{sizeof\_code}|": 3584,
      "|\textbf{sizeof\_headers}|": 1024,
      "|\textbf{sizeof\_heap\_commit}|": 4096
    }
  },
  "|\textbf{imports}|": {
    "KERNEL32.dll": [ "GetTickCount" ],
    ...
  },
  "|\textbf{exports}|": []
  "|\textbf{section}|": {
    "|\textbf{entry}|": ".text",
    "|\textbf{sections}|": [
      {
        "|\textbf{name}|": ".text",
        "|\textbf{size}|": 3584,
        "|\textbf{entropy}|": 6.368472139761825,
        "|\textbf{vsize}|": 3270,
        "|\textbf{props}|": [ "CNT_CODE", "MEM_EXECUTE", "MEM_READ"]
      },
      ...
    ]
  },
  "|\textbf{histogram}|": [ 3818, 155, ..., 377 ],
  "|\textbf{byteentropy}|": [0, 0, ... 2943 ],
  "|\textbf{strings}|": {
    "|\textbf{numstrings}|": 170,
    "|\textbf{avlength}|": 8.170588235294117,
    "|\textbf{printabledist}|": [ 15, ... 6 ],
    "|\textbf{printables}|": 1389,
    "|\textbf{entropy}|": 6.259255409240723,
    "|\textbf{paths}|": 0,
    "|\textbf{urls}|": 0,
    "|\textbf{registry}|": 0,
    "|\textbf{MZ}|": 1
  },
}    
\end{lstlisting}
\caption{Raw features extracted from a single PE file.\label{fig:feature_example}}
\end{figure}

\begin{figure}[t]
\centering
\includegraphics[scale=0.6]{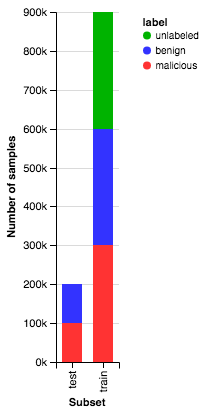}
\caption{Distribution of malicious, benign and unlabeled samples in the training and test sets\label{fig:split}}
\end{figure}

\begin{figure}[t]
\centering
\includegraphics[scale=0.6]{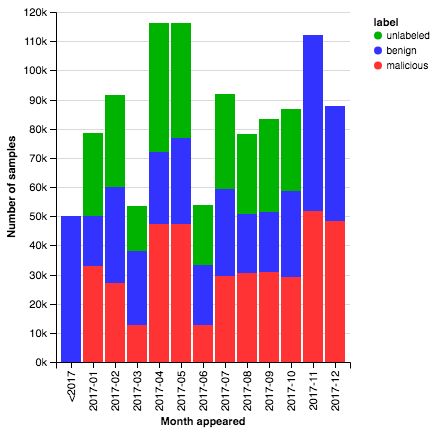}
\caption{A temporal distribution of the dataset, available from chronology data available in the metadata, with \texttt{2017-11} and \texttt{2017-12} corresponding to the test set\label{fig:temporal}}
\end{figure}

In crafting the EMBER dataset, we considered several practical use cases and research studies, including the following.
\begin{itemize}
    \item Compare machine learning models for malware detection.
    \item Quantify model degradation and concept drift over time.
    \item Research interpretable machine learning.
    \item Compare features for malware classification, particularly novel features not represented in the EMBER dataset.  This requires an extensible dataset.
    \item Compare to featureless end-to-end deep learning.  This may require code to extract features from a new dataset, or shas256 hashes to build a raw binary dataset to match EMBER.
    \item Research adversarial attacks against machine learning malware, and subsequent defense strategies.
    \item Leverage unlabeled samples via unsupervised learning for PE file representation or semi-supervised learning for classification.
\end{itemize}
Considerations of these use cases led to the data structure outlined in this section.

\subsection{Data layout}
The EMBER dataset consists of a collection of JSON lines files, where each line contains a single JSON object.  Each object includes the following types in data:
\begin{itemize}
    \item the sha256 hash of the original file as a unique identifier;
    \item coarse time information (month resolution) that establishes an estimate of when the file was first seen;
    \item a label, which may be 0 for benign, 1 for malicious or -1 for unlabeled; and
    \item eight groups of raw features that include both parsed values as well as format-agnostic histograms.
\end{itemize}
Details of each feature type are described in more detail below, and an example is shown in Figure \ref{fig:feature_example}.  

For convenience our dataset is comprised of raw features that are human readable.  We provide code that produces from raw features a numeric feature vector required for model building.  This allows researchers to decouple raw features from the vectorizing strategies.  In our code, we provide a default method that produces a feature matrix for training a baseline model, and should be suitable for most use cases. However, the availability of raw features may allow studies into explainable machine learning, or feature importance as in \cite{shafiq2009framework}.  We have also included unlabeled samples in the training set to encourage research in semi-supervised learning approaches (see Figure \ref{fig:split}), which appears to be a relatively unexplored area for malware classification in published literature. As another consideration, we temporally split the training/test sets (see Figure \ref{fig:temporal}) to mimic generational dependencies of both malicious and benign software.  The coarse time stamps for one year of malicious and benign files may also allow for simple longitudinal studies.  Including the sha256 hash of the original file allows researchers to link features to the raw binaries, including other metadata that may be available through file sharing sites like VirusShare or VirusTotal \cite{virusshare,virustotal}.  For convenience, we ensured the files labeled benign in EMBER were available in VirusTotal and at the time of collection, no vendors detected them as malicious. Likewise, we ensured that files labeled malicious in EMBER were available in VirusTotal and had more than 40 vendors report as malicious. As such, EMBER is a relatively ``easy'' dataset. 

\subsection{Feature set description}
The EMBER dataset consists of eight groups of raw features that include both parsed features and format-agnostic histograms and counts of strings.  In what follows, we make a distinction between \textit{raw features} (the dataset provided) and \textit{model features} (or \texttt{vectorized features}) derived from the dataset.  \textit{Model features} represent a feature matrix of fixed size used for training a model, representing a numerical summary of the raw features, wherein strings, imported names, exported names, etc., are captured using the feature hashing trick \cite{weinberger2009feature}.  The feature matrix is not explicitly provided in the published dataset, but code is provided to convert raw features to model features to train a baseline model.  For convenience, we use the implementation provided by \texttt{scikit-learn} \cite{scikit-learn}.  Where appropriate, in the feature descriptions below, we note the number of bins used for the feature hashing trick.

\subsubsection{Parsed features}
The dataset includes five groups of features that are extracted after parsing the PE file.  We leverage the Library to Instrument Executable Formats \cite{LIEF} as a convenient PE parser.  LIEF names are used for strings that represent symbolic objects, such as characteristics and properties.  For some examples of these strings, the reader is referred to Figure \ref{fig:feature_example}.  Each of the parsed feature types are described in more detail below.

\paragraph{General file information.} The set of features in the general file information group includes the file size and basic information obtained from the PE header: the virtual size of the file, the number of imported and exported functions, whether the file has a debug section, thread local storage, resources, relocations, or a signature, and the number of symbols.

\paragraph{Header information.}  From the COFF header, we report the timestamp in the header, the target machine (string) and a list of image characteristics (list of strings).  From the optional header, we provide the target subsystem (string), DLL characteristics (a list of strings), the file magic as a string (e.g., ``PE32''), major and minor image versions, linker versions, system versions and subsystem versions, and the code, headers and commit sizes.  To create model features, string descriptors such as DLL characteristics, target machine, subsystem, etc. are summarized using the feature hashing trick prior to training a model, with 10 bins allotted for each noisy indicator vector.

\paragraph{Imported functions.}  We parse the import address table and report the imported functions by library.  To create model features for the baseline model, we simply collect the set of unique libraries and use the hashing trick to sketch the set (256 bins).  Similarly, we use the hashing trick (1024 bins) to capture individual functions, by representing each as a string such as \texttt{library:FunctionName} pair (e.g., \texttt{kernel32.dll:CreateFileMappingA}).

\paragraph{Exported functions.}  The raw features include a list of the exported functions.  These strings are summarized into model features using the hashing trick with 128 bins.

\paragraph{Section information.} Properties of each section are provided and include the  name, size, entropy, virtual size, and a list of strings representing section characteristics.  The entry point is specified by name.  To convert to model features, we use the hashing trick on (section name, value) pairs to create vectors containing section size, section entropy, and virtual size (50 bins each). We also use the hashing trick to capture the characteristics (list of strings) for the entry point.

\subsubsection{Format-agnostic features}
The EMBER dataset also includes three groups of features that are format agnostic, in that they do not require parsing of the PE file for extraction: a raw byte histogram, byte entropy histogram based on work previously published in \cite{saxe2015deep}, and string extraction. 

\paragraph{Byte histogram.}  The byte histogram contains 256 integer values, representing the counts of each byte value within the file.  When generating model features, this byte histogram is normalized to a distribution, since the file size is represented as a feature in the general file information.

\paragraph{Byte-entropy histogram.} The byte entropy histogram approximates the joint distribution $p(H,X)$ of entropy $H$ and byte value $X$.  This is done as described in \cite{saxe2015deep}, by computing the scalar entropy $H$ for a fixed-length window and pairing it with each byte occurrence within the window.  This is repeated as the window slides across the input bytes. In our implementation, we use a window size of 2048 and a step size of 1024 bytes, with $16 \times 16$ bins that quantize entropy and the byte value.  Before training, we normalize these counts to sum to unity.

\paragraph{String information.} The dataset includes simple statistics about printable strings (consisting of characters in the range 0x20 to 0x7f, inclusive) that are at least five printable characters long.  In particular, reported are the number of strings, their average length, a histogram of the printable characters within those strings, and the entropy of characters across all printable strings.  The printable characters distribution provides distinct information from the byte histogram information above since it is derived only from strings containing at least five consecutive printable characters.  In addition, the string feature group includes the number of strings that begin with \texttt{C:\textbackslash} (case insensitive) that may indicate a path, the number of occurrences of \texttt{http://} or \texttt{https://} (case insensitive) that may indicate a URL, the number of occurrences of \texttt{HKEY\textunderscore} that may indicate a registry key, and the number of occurrences of the short string \texttt{MZ} that may provide weak evidence of a Windows PE dropper or bundled executables.  By providing a simple statistical summary of strings rather than a listing of raw strings, we mitigate privacy concerns that may exist for some benign files.

\section{Experiments}
\label{sec:results}
\begin{figure}[t]
\centering
\includegraphics[width=3.8in]{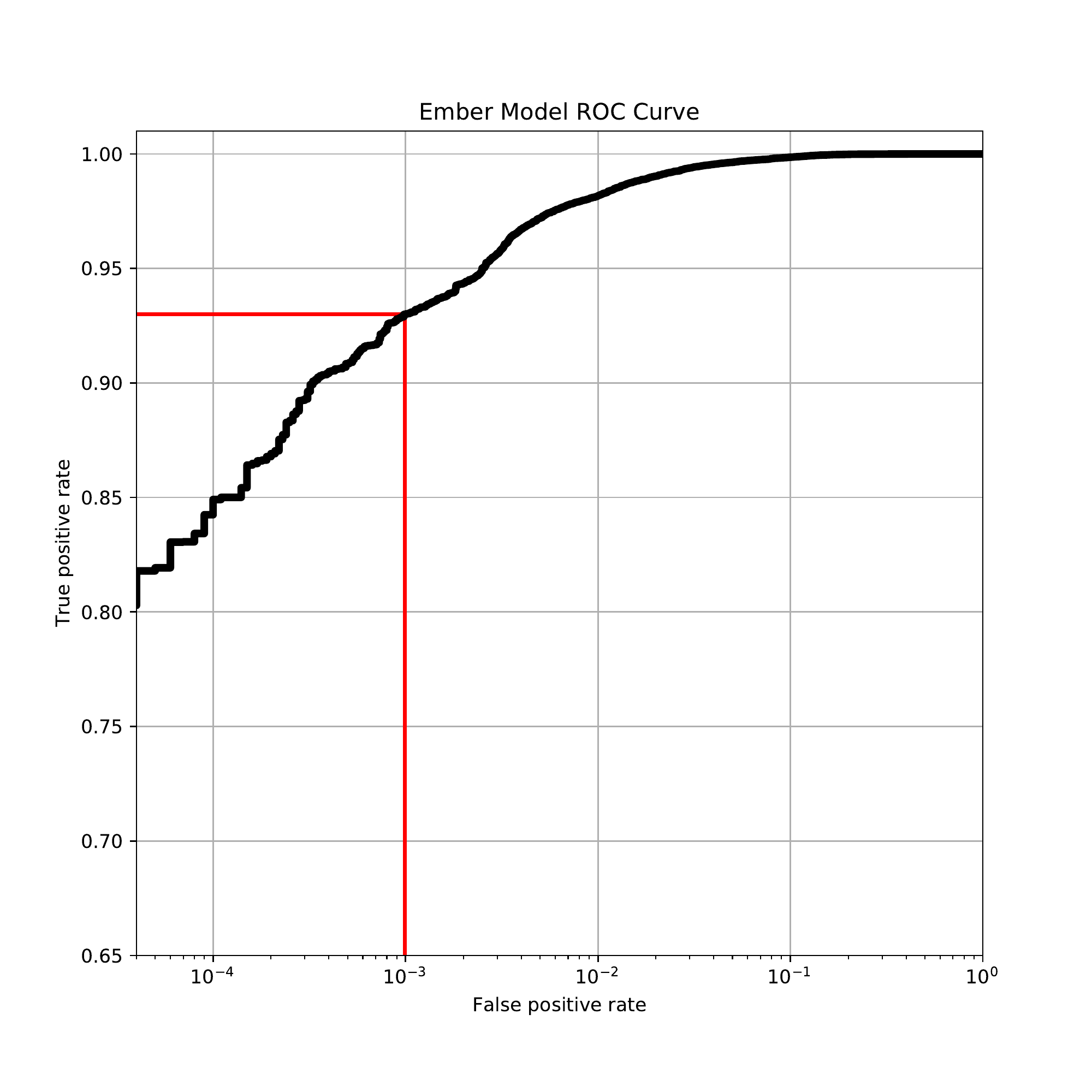}
\caption{ROC curve with log scale for false positive rate (FPR). The threshold shown (red) corresponds to a 0.1\% FPR and a detection rate about 93\%.  At 1\% FPR the detection rate exceeds 98\%.\label{fig:roc}}
\end{figure}

\begin{figure}[t]
\centering
\includegraphics[width=3.8in]{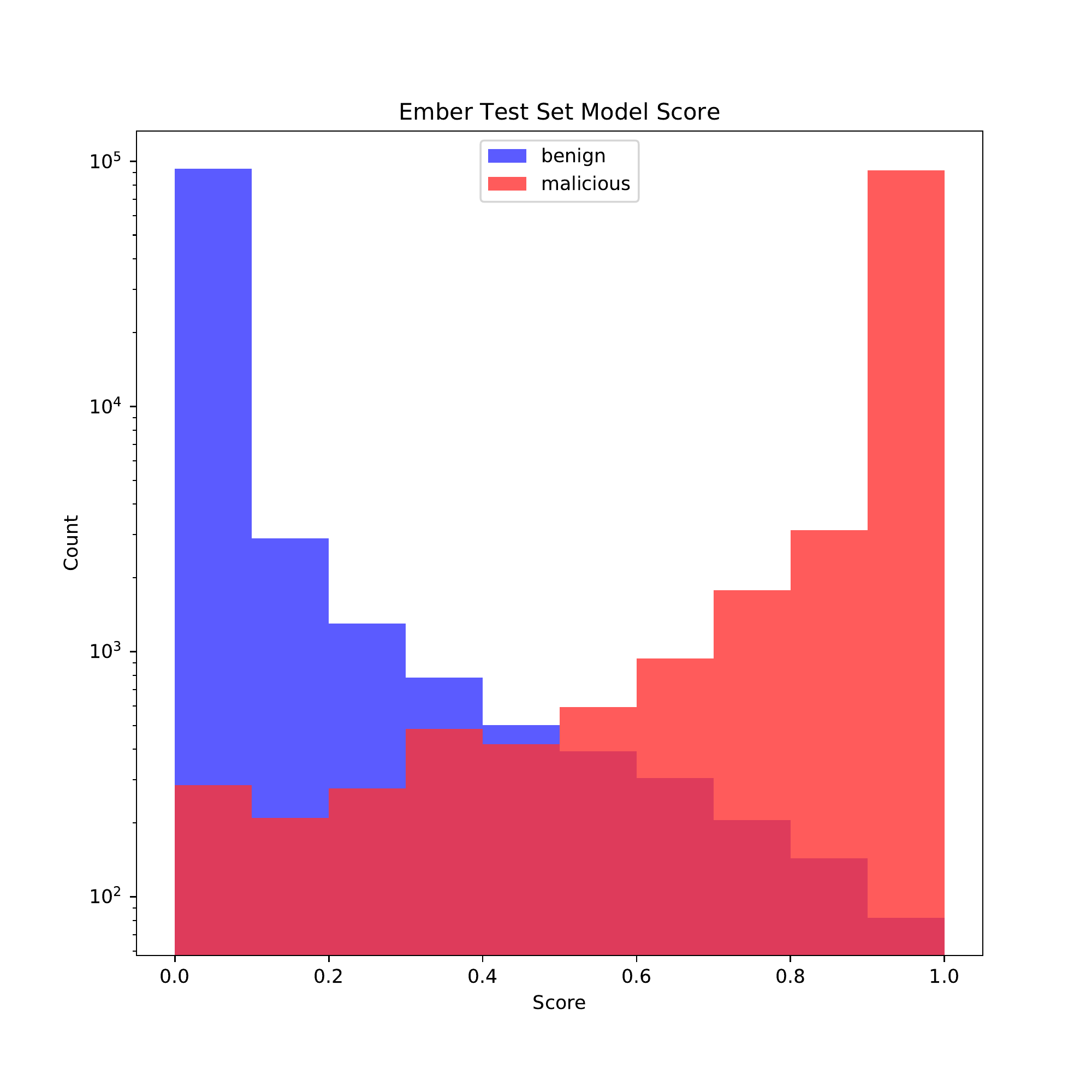}
\caption{Distribution of model test scores on the test set (note the logarithmic scale)\label{fig:scores}}
\end{figure}

EMBER includes code that demonstrates how to use the raw features in the training set (labeled samples only) for building a supervised learning model, which we provide as a baseline model. The model building process consists of vectorizing the raw features (each object into a vector of dimension 2351), using the feature hashing trick where necessary, as described previously.  On a 2015 MacBook Pro i7, it took 20 hours to vectorize the raw features into model features.  From the vectorized features, we trained a gradient-boosed decision tree (GBDT) model using \texttt{LightGBM} with default parameters (100 trees, 31 leaves per tree), resulting in fewer than 10K tunable parameters \cite{ke2017lightgbm}. Model training took 3 hours. Baseline model performance may be much improved with appropriate hyper-parameter optimization, which is of less interest to us in this work.

A ROC curve of the resulting model is shown in Figure \ref{fig:roc}, and a distribution of scores for malicious and benign samples in the test set is shown in Figure \ref{fig:scores}.  The ROC AUC exceeds 0.99911.  A threshold of 0.871
on the model score results in less than 0.1\% FP rate at a detection rate exceeding 92.99\%.  At less than 1\% FP rate, the model exceeds 98.2\% detection rate.

As discussed in Section \ref{sec:background}, it has been suggested that owing to the dataset it was trained on, the Adobe Malware Classifier is biased towards non-Windows vs. Windows classification, rather than a true malicious vs. benign problem \cite{seymour2016howto}.  We evaluated the pre-trained J48 model on our test set and found that it exhibits an alarming 53\% false positive rate and an 8\% false negative rate.  This would seem to substantiate previous claims about dataset bias. However, whether this poor performance can be ascribed to stale training data or dataset bias or both is out of the scope of this paper.  But clearly, it is an inappropriate baseline model.

As a comparative study, we trained MalConv \cite{raff2017malware} on the raw binaries underlying the dataset.  We used the model architecture and training setup as prescribed and verified by the authors, except that we train with a batch size of 100 instead of 256 due to GPU memory constraints.  We trained using data parallelism across two Titan X (Pascal) GPUs.  Each epoch took 25 hours, and we trained for 10 epochs (10 days).  The resulting model has roughly 1M parameters.  Applied to the raw binaries corresponding to the EMBER test set, the Malconv ROC AUC is 0.99821, corresponding to a 92.2\% detection rate at a false positive rate less than 0.1\%, or a 97.3\% detection rate at a less than 1\% false positive rate.  This is slightly lower performance than using \texttt{LightGBM} with no hyper-parameter tuning.  Evidently, despite increased model size and computational burden, featureless deep learning models have yet to eclipse the performance of models that leverage domain knowledge via parsed features.

\section{Discussion}
\label{sec:conclusion}
To our knowledge, the EMBER dataset represents the first large public dataset for machine learning malware detection (which must include benign files).  It is the authors' hope that the dataset is useful to spur innovation in machine learning malware detection.  We considered a number of research use cases in Section \ref{sec:method} including comparing model performance, adversarial machine learning offense and defense, semi-supervised learning for malware detection, and many more research areas.

With the dataset, we have also released a simple non-optimized benchmark \texttt{LightGBM} model.  Simple approaches to immediately improve model performance include feature selection to eliminate noisy features and hyper-parameter optimization via grid search. Nevertheless, we demonstrate that the out-of-the-box \texttt{LightGBM} model trained on these features outperforms recently published work in end-to-end deep learning for malware detection \cite{raff2017malware}.  Thus, in addition to a benchmark dataset, we hope that EMBER can provide a simple means for benchmarking model performance of novel architectures including end-to-end deep learning.

The dataset and source code are available at \SOURCECODE.

\pagebreak[4]
\section*{Acknowledgements}
The authors wish to thank Peter Silberman for his careful review of the code repository and dataset, with useful suggestions for improvement.

\bibliographystyle{abbrv}
\bibliography{ref}
\end{document}